\documentstyle[12pt]{article}
\begin{document}
\parskip=3pt
\baselineskip=22pt
\centerline{\Large\bf Generalized $t$-$j$ Model
\footnote{\sf Supported in part by the National Natural Science Foundation
of China.}}
\vspace{4ex}
\centerline{\large\sf  Shao-Ming Fei ~~and ~~Rui-Hong Yue}
\vspace{3ex}
\centerline{\sf         CCAST (World Laboratory)}
\centerline{\sf         P. O. Box 8730,         Beijing 100080, China}
\vspace{2ex}
\centerline{\sf and}
\vspace{2ex}
\centerline{\sf Institute of Theoretical Physics, Academia Sinica}
\centerline{\sf P. O. Box 2735, Beijing 100080, China\footnote{\sf
Mailing address}}
\vspace{4ex}

\begin{center}
\begin{minipage}{5in}
\centerline{\large\bf   Abstract}
\vspace{3ex}

{\sf ~~~~~By parameterizing the t-j model we present a new electron
correlation model with one free parameter for high-temperature
superconductivity. This model is of $SU_{q}(1,2)$ symmetry. The energy
spectrums are shown to be modulated by the free parameter in the model. The
solution and symmetric structures of the Hilbert space, as well as the
Bethe ansatz approach are discussed for special cases.}

\end{minipage}
\end{center}
\newpage

Strongly correlated electronic systems are believed to be important in
studying the
phenomenon of high-temperature superconductivity \cite{and88,fuku89}.
An appropriate starting model suggested by Anderson is the t-j model
\cite{and87,zhang88}. The model describes the behavior of electrons
on a discrete lattice  with Hamiltonian including nearest-neighbour hopping
(t) and antiferromagnetic exchange (j). And the Hilbert space admits no
double occupancy of any single site. As the two dimensional systems may share
features of one dimensional systems\cite{and90}, the t-j model in one dimension
has been extensively investigated. The model is shown to be integrable and
supersymmetric when $j=\pm 2t$ \cite{korepin,sarkar}.
Nevertheless, the phenomenon of
high-temperature superconductivity greatly depends on detailed material.
While in supersymmetric t-j model there is no non-trivial free parameters
left. By taking into account some physical considerations, in this letter
we present a generalized t-j model. The interest is that this model
degenerates into a integrable one with a free parameter $q$ and q-deformed
supersymmetric symmetry.

Electrons on a lattice are described by canonical Fermi operators
$c^{+}_{j\sigma}$ and $c_{j\sigma}$ satisfying anti-commutation relations
given by
\begin{equation}
\{c^{+}_{i\sigma},c_{j\sigma^{'}}\}=\delta_{ij}\delta_{\sigma\sigma^{'}}~,
\end{equation}
where $\sigma=\uparrow,\downarrow$; $j=1,\cdots,L$ and $L$ is the total
number of lattice sites. $c_{i\sigma}$ annihilates an electron of spin
$\sigma$ at site $i$. The Fock vacuum $\vert 0\rangle$ satisfies
$c_{i\sigma}\vert 0\rangle=0$. As the double occupancy is not allowed,
there are three possible electronic states at a given lattice site $i$:
$$
\vert 0\rangle,~~~\vert +\rangle\equiv c^{+}_{j\uparrow}\vert 0\rangle
=\vert\uparrow\rangle,~~~
\vert -\rangle\equiv c^{+}_{j\downarrow}\vert 0\rangle
=\vert\downarrow\rangle~.
$$

The Hamiltonian of supersymmetric t-j model on a lattice of L sites is
given by the following expression:
\begin{equation}\label{tj}
\begin{array}{rcl}
H_{tj}&=&\displaystyle\sum^{L-1}_{i=1}\left[X^{+-}_{i}X^{-+}_{i+1}
+X^{-+}_{i}X^{+-}_{i+1}-X^{-0}_{i}X^{0-}_{i+1}+X^{0-}_{i}X^{-0}_{i+1}
-X^{+0}_{i}X^{0+}_{i+1}\right.\\[5mm]
&&\left.+X^{0+}_{i}X^{+0}_{i+1}+n_{i}^{+}n_{i+1}^{+}
+n_{i}^{-}n_{i+1}^{-}-n_{i}^{0}n_{i+1}^{0}\right]~,
\end{array}
\end{equation}
where
\begin{equation}\label{x}
X^{\alpha\beta}_{i}=\vert\alpha\rangle_{i}\,{}_{i}\langle\beta\vert~,~~~~
\alpha,~\beta=0,+,-,
\end{equation}
are the local generators of the supersymmetric
algebra $SU(1,2)$ and
\begin{equation}\label{n}
n_{i}^{0}=\vert 0\rangle_{i}\,{}_{i}\langle 0\vert~,~~~
n_{i}^{+}=\vert +\rangle_{i}\,{}_{i}\langle +\vert~,~~~
n_{i}^{-}=\vert -\rangle_{i}\,{}_{i}\langle -\vert
\end{equation}
are the number operators
of holes, spin-up electrons and spin-down electrons at site $i$
respectively. It is direct to prove that $H_{tj}$ commutes with the
total operators of $SU(1,2)$ on lattice.

In parameterizing the usual supersymmetric t-j Hamiltonian (\ref{tj}), we
reasonably distinguish the interactions between spin up(down) electrons
and between holes. Therefore the coupling constants of
$n_{i}^{+}n_{i+1}^{+}$ and $n_{i}^{-}n_{i+1}^{-}$ are different to the one
of $n_{i}^{0}n_{i+1}^{0}$.

We also suppose that the chemical potentials of spin up(down) electrons and
holes are different. The terms for chemical potential is then of the form
\begin{equation}
\mu=\sum_{i=1}^{L}c\,n_{i}^{0}+c^{'}\,(n_{i}^{+}+n_{i}^{-})~,
\end{equation}
where $c$ and $c^{'}$ are chemical potentials of a hole and an electron
respectively. Accounting to the no double occupancy
condition,
\begin{equation}\label{condition}
n_{i}^{0}+n_{i}^{+}+n_{i}^{-}=1,
\end{equation}
the chemical potential simply becomes
\begin{equation}\label{mu}
\mu=\displaystyle\sum_{i}^{L}n_{i}^{0}\,(c-c^{'})
\end{equation}
modular a constant term.

In addition we consider the nearest-neighbour interactions between
electrons with opposed spin direction. For an electron $c^{+}_{j\sigma}$
located at site $j$, the general form of the interaction is
$n^{\bar{\sigma}}_{j-1}n^{\sigma}_{j}+n^{\sigma}_{j}n^{\bar{\sigma}}_{j+1}$,
where ${\sigma}$ takes $\uparrow$ and $\bar{\sigma}$ takes $\downarrow$
and vice versa.
If we impose the saturation condition on it, the interaction only
contributes one term to the Hamiltonian. Without loss of generality, we
can assume that the interaction between opposed direction spin electrons is
\begin{equation}
\sum_{i=1}^{L-1}n_{i}^{-}n_{i+1}^{+}~,
\end{equation}
which may be in a sense considered as the
first order expansion of Cooper pair in coordinate space. In the following,
we will see that this term plays an important role in resumption of
symmetry of the system.

Based on the analysis above and accounting to suitable boundary conditions,
we present a modified t-j Hamiltonian with parameters:
\begin{equation}\label{hp}
\begin{array}{rcl}
H&=&\displaystyle\sum^{L-1}_{i=1}\left[X^{+-}_{i}X^{-+}_{i+1}
+X^{-+}_{i}X^{+-}_{i+1}-X^{-0}_{i}X^{0-}_{i+1}+X^{0-}_{i}X^{-0}_{i+1}
-X^{+0}_{i}X^{0+}_{i+1}\right.\\[5mm]
&&\left.+X^{0+}_{i}X^{+0}_{i+1}+\lambda n_{i}^{-}n_{i+1}^{+}
+\gamma (n_{i}^{+}n_{i+1}^{+}+n_{i}^{-}n_{i+1}^{-})
+\theta n_{i}^{0}n_{i+1}^{0}+\tau n_{i}^{0}\right]~,
\end{array}
\end{equation}
where $\lambda,\gamma,\theta,\tau$ are free parameters. This Hamiltonian
no longer possesses the $SU(1,2)$ symmetry and is not integrable in general.
To recover the broken symmetry in some contents we note that the generators
$X^{\alpha\beta}_{i}$, $\alpha,~\beta=0,+,-$ in (\ref{x})
are also the trivial representations of the q-deformed algebra
$SU_{q}(1,2)$. Therefore a reasonable candidate for symmetry imposed
on the system is $SU_{q}(2)$.
Before studying the symmetry of the system, we give some
preliminary knowledge of $SU_{q}(1,2)$. This algebra is spanned by generators
$X^{\alpha\beta}~,\alpha,\beta=+,-,0$. They satisfy the following algebraic
relations:
\begin{equation}\label{algebra}
\begin{array}{ll}
[X^{+-},X^{-+}]=[X^{++}-X^{--}]_{q}~,&\\[3mm]
\{X^{-0},X^{0-}\}=[X^{--}+X^{00}]_{q}~,&\\[3mm]
[X^{++},X^{+-}]=X^{+-}~,~~~&
[X^{++},X^{+0}]=X^{+0}\\[3mm]
[X^{++},X^{-+}]=-X^{-+}~,~~~&
[X^{++},X^{0+}]=-X^{0+}\\[3mm]
[X^{--},X^{-+}]=X^{-+}~,~~~&
[X^{--},X^{-0}]=X^{-0}\\[3mm]
[X^{--},X^{+-}]=-X^{+-}~,~~~&
[X^{--},X^{0-}]=-X^{0-}\\[3mm]
[X^{00},X^{0+}]=X^{0+}~,~~~&
[X^{00},X^{0-}]=X^{0-}\\[3mm]
[X^{00},X^{+0}]=-X^{+0}~,~~~&
[X^{00},X^{-0}]=-X^{-0}\\[3mm]
(X^{-0})^{2}=0~,~~~&(X^{0-})^{2}=0\\[3mm]
\end{array}
\end{equation}
and the Sierre relations
\begin{equation}\label{sirre}
\begin{array}{l}
(X^{+-})^{2}X^{-0}-(q+q^{-1})X^{+-}X^{-0}
X^{+-}+X^{-0}(X^{+-})^{2}=0~,\\[3mm]
(X^{-+})^{2}X^{0-}-(q+q^{-1})X^{-+}X^{0-}
X^{-+}+X^{0-}(X^{-+})^{2}=0~,
\end{array}
\end{equation}
where
$$
[X]_{q}\equiv \frac{q^{x}-q^{-x}}{q-q^{-1}}~.
$$

As a quantum algebra, $SU_{q}(1,2)$ has non-trivial Hopf algebraic
structures with operations coproduct, counit and antipode. Here we only
give the coproduct expressions:
\begin{equation}\label{corep}
\begin{array}{rcl}
\Delta(X^{\alpha\alpha})&=&X^{\alpha\alpha}\otimes {\bf 1}+
{\bf 1}\otimes X^{\alpha\alpha}~,~~~\alpha=+,-,0\\[3mm]
\Delta(X^{-+})&=&
X^{-+}\otimes q^{(X^{++}-X^{--})/2}+q^{-(X^{++}-X^{--})/2}\otimes X^{-+}\\[3mm]
&=&X^{-+}\otimes\{1+(q^{\frac{1}{2}}-1)X^{++}+(q^{-\frac{1}{2}}-1)X^{--}\}
\\[3mm]
&&+\{1+(q^{-\frac{1}{2}}-1)X^{++}+(q^{\frac{1}{2}}-1)X^{--}\}\otimes X^{-+}
{}~,\\[3mm]
\Delta(X^{+-})&=&
X^{+-}\otimes q^{(X^{++}-X^{--})/2}+q^{-(X^{++}-X^{--})/2}\otimes
X^{+-}\\[3mm]
&=&
X^{+-}\otimes\{1+(q^{\frac{1}{2}}-1)X^{++}+(q^{-\frac{1}{2}}-1)X^{--}\}\\[3mm]
&&+\{1+(q^{-\frac{1}{2}}-1)X^{++}+(q^{\frac{1}{2}}-1)X^{--}\}\otimes X^{+-}
{}~,\\[3mm]
\Delta(X^{0-})&=&
X^{0-}\otimes q^{(X^{00}+X^{--})/2}+q^{-(X^{00}+X^{--})/2}\otimes
X^{0-}\\[3mm]
&=&
X^{0-}\otimes\{1+(q^{\frac{1}{2}}-1)(X^{--}+X^{00})\}\\[3mm]
&&+\{1+(q^{-\frac{1}{2}}-1)(X^{--}+X^{00})\}\otimes X^{0-}
{}~,\\[3mm]
\Delta(X^{-0})&=&
X^{-0}\otimes q^{(X^{00}+X^{--})/2}+q^{-(X^{00}+X^{--})/2}\otimes
X^{-0}\\[3mm]
&=&
X^{-0}\otimes\{1+(q^{\frac{1}{2}}-1)(X^{--}+X^{00})\}\\[3mm]
&&+\{1+(q^{-\frac{1}{2}}-1)(X^{--}+X^{00})\}\otimes X^{-0}
{}~,\\[3mm]
\Delta(X^{0+})&=&
X^{0+}\otimes q^{(X^{++}+X^{00})/2}+q^{-(X^{++}+X^{00})/2}\otimes
X^{-+}\\[3mm]
&&+(q-q^{-1})q^{-(X^{++}-X^{--})/2}X^{0-}\otimes
q^{(X^{--}+X^{00})/2}X^{-+}\\[3mm]
&=&
X^{0+}\otimes\{1+(q^{\frac{1}{2}}-1)(X^{++}+X^{00})\}\\[3mm]
&&+\{1+(q^{-\frac{1}{2}}-1)(X^{++}+X^{00})\}\otimes X^{0+}
+(q-q^{-1})X^{0-}\otimes X^{-+}
{}~,\\[3mm]
\Delta(X^{+0})&=&
X^{+0}\otimes q^{(X^{++}+X^{00})/2}+q^{-(X^{++}+X^{00})/2}\otimes
X^{+0}\\[3mm]
&&+(q-q^{-1})q^{-(X^{++}-X^{--})/2}X^{-0}\otimes
q^{(X^{--}+X^{00})/2}X^{+-}\\[3mm]
&=&
X^{+0}\otimes\{1+(q^{\frac{1}{2}}-1)(X^{++}+X^{00})\}\\[3mm]
&&+\{1+(q^{-\frac{1}{2}}-1)(X^{++}+X^{00})\}\otimes X^{+0}
+(q-q^{-1})X^{-0}\otimes X^{+-}~.
\end{array}
\end{equation}

The co-product operator $\Delta$ are algebraic isomorphism,
$\Delta(ab)=\Delta(a)\Delta(b)~,~~\forall a,b\in U_{q}(1,2)$.
$X^{0+}$ and $X^{+0}$ are the algebra elements in the sense that
\begin{equation}
\begin{array}{l}
X^{+0}=q^{\frac{1}{2}}X^{+-}X^{-0}-q^{-\frac{1}{2}}X^{-0}X^{-+}~,\\[3mm]
X^{0+}=q^{-\frac{1}{2}}X^{0-}X^{-+}-q^{\frac{1}{2}}X^{-+}X^{0-}~.
\end{array}
\end{equation}
Their co-product representations $\Delta(X^{+0})$ and $\Delta(X^{0+})$
are obtained by the actions of co-product operator $\Delta$. It is not
difficult to show that operators defined by equation (\ref{corep}) satisfy
the same relations (\ref{algebra}) as $X^{\alpha\beta}$.

Now we return to study what conditions will be imposed on the parameters
appearing in Hamiltonian (\ref{hp}), when the system enjoys the symmetry of
quantum group $SU_{q}(1,2)$. Define
\begin{equation}
{\cal X}^{\alpha\beta}=\Delta^{L-1}(X^{\alpha\beta})
=(\Delta\otimes id)\Delta^{L-2}(X^{\alpha\beta})~.
\end{equation}
Explicitly
\begin{equation}\label{xtotal}
\begin{array}{l}
{\cal X}^{\alpha\alpha}=\displaystyle\sum_{i=1}^{L}{\bf 1}
\otimes\cdots\otimes{\bf 1}
\otimes X^{\alpha\alpha}\otimes{\bf 1}\otimes \cdots\otimes {\bf 1}
{}~,~~~\alpha=+,-,0\\[5mm]
{\cal X}^{\bar{\alpha}\alpha}=\displaystyle\sum_{i=1}^{L}q^{-H_{1}/2}
\otimes\cdots\otimes q^{-H_{1}/2}
\otimes X^{\bar{\alpha}\alpha}\otimes q^{H_{1}/2}
\cdots\otimes q^{H_{1}/2}~,~~~\alpha,\bar{\alpha}=+,-\\[5mm]
{\cal X}^{0-}=\displaystyle\sum_{i=1}^{L}q^{-H_{2}/2}
\otimes\cdots\otimes q^{-H_{2}/2}
\otimes X^{0-}\otimes q^{H_{2}/2}
\cdots\otimes q^{H_{2}/2}~,\\[5mm]
{\cal X}^{-0}=\displaystyle\sum_{i=1}^{L}q^{-H_{2}/2}
\otimes\cdots\otimes q^{-H_{2}/2}
\otimes X^{-0}\otimes q^{H_{2}/2}
\cdots\otimes q^{H_{2}/2}~,
\end{array}
\end{equation}
where $H_{1}=X^{++}-X^{--}$ and $H_{2}=X^{00}+X^{--}$.
The operator ${\cal X}^{0+}$ (${\cal X}^{+0}$) can also be represented in
terms of $X^{-+}$ and $X^{0-}$ ($X^{+-}$ and $X^{-0}$). These operators
satisfy the $SU_{q}(1,2)$ algebraic relations (\ref{algebra}).

It is easy to show that the bosonic operators ${\cal X}^{\alpha\alpha}$
commute with the Hamiltonian (\ref{hp}) for arbitrary parameters
$\lambda$, $\gamma$, $\theta$ and $\tau$. Therefore we only need to
calculate
\begin{equation}
\begin{array}{l}
[{\cal X}^{+-},H]\\[4mm]
=\displaystyle\sum_{j=1}^{L-1}
q^{-H_{1}/2}\otimes\cdots\otimes q^{-H_{1}/2}\otimes
[q^{-H_{1}/2}\otimes X^{+-}_{j+1},H_{j\,j+1}]
\otimes q^{H_{1}/2}\otimes\cdots\otimes q^{H_{1}/2}\\[6mm]
+q^{-H_{1}/2}\otimes\cdots\otimes q^{-H_{1}/2}\otimes
[X^{+-}_{j}\otimes q^{H_{1}/2},H_{j\,j+1}]
\otimes q^{H_{1}/2}\otimes\cdots\otimes q^{H_{1}/2}\\[5mm]
=\displaystyle\sum_{j=1}^{L-1}(q^{1/2}-q^{-1/2}\gamma)X_{j}^{++}X_{j+1}^{+-}
+(q^{1/2}\lambda-q^{-1/2}+q^{1/2}\gamma)X_{j}^{--}X_{j+1}^{+-}\\[6mm]
+(q^{-1/2}-q^{1/2}\gamma+q^{1/2}\lambda)X^{+-}_{j}X_{j+1}^{++}
+(-q^{1/2}+q^{-1/2}\gamma)X_{j}^{+-}X_{j+1}^{--}
\end{array}
\end{equation}

Then the following conditions are necessary for $[{\cal X}^{+-},H]=0$
\begin{equation}
\left\{ \begin{array}{l}\gamma=q\\[2mm]\lambda=q-q^{-1}\end{array}\right.
\end{equation}
Similarly, the vanishing condition for commutator $[{\cal X}^{0+},H]=0$
imposes the conditions $\tau=\lambda$ and $\theta=-\gamma$. Taking
$\lambda=\tau=q-q^{-1}$ and $\gamma=-\theta=q$ in (\ref{hp}) we get a
parameterized Hamiltonian with q-deformed symmetry $SU_{q}(1,2)$,
\begin{equation}\label{qtj}
\begin{array}{rcl}
H_{tj}^{q}&=&\displaystyle\sum^{L-1}_{i=1}\left[X^{+-}_{i}X^{-+}_{i+1}
+X^{-+}_{i}X^{+-}_{i+1}-X^{-0}_{i}X^{0-}_{i+1}+X^{0-}_{i}X^{-0}_{i+1}
-X^{+0}_{i}X^{0+}_{i+1}+X^{0+}_{i}X^{+0}_{i+1}\right.\\[5mm]
&&\left.+q(n_{i}^{+}n_{i+1}^{+}+n_{i}^{-}n_{i+1}^{-}-n_{i}^{0}n_{i+1}^{0})
+(q-q^{-1})(n_{i}^{-}n_{i+1}^{+}+n_{i}^{0})\right]~.
\end{array}
\end{equation}
The calculations to check that the other operators commutes with $H_{tj}^{q}$
is tedious but direct.
Therefore the modified t-j Hamiltonian (\ref{qtj}) is of $SU_{q}(1,2)$
symmetry. It has a free parameter $q$. When $q$ approaches one it becomes
the usual t-j Hamiltonian (\ref{tj}).

This generalized t-j model is again integrable. Similar to the case of
supersymmetric t-j model \cite{korepin}, it can be exactly solved in terms
of the algebraic Bethe ansatz. The symmetry algebra is now $SU_{q}(1,2)$.
Moreover, the usual coordinate Bethe ansatz also gives complete energy
spectrums and eigenstates of the generalized model by using the symmetry
algebra operators. The detailed exact solutions and phase diagram analysis
are to appear in our paper \cite{we}. In this letter we calculate the model
at a few lattice sites, so as to show the symmetric structures of the
Hilbert space and the roles played by the free parameter $q$.

First we consider the case $L=2$. The Hamiltonian is simply
\begin{equation}
\begin{array}{rcl}
H&=&X^{+-}\otimes X^{-+}+X^{-+}\otimes X^{+-}-X^{-0}\otimes X^{0-}\\[3mm]
&&+X^{0-}\otimes X^{-0}-X^{+0}\otimes X^{0+}
+X^{0+}\otimes X^{+0}+q(n^{+}\otimes n^{+}\\[3mm]
&&+n^{-}\otimes n^{-}-n^{0}\otimes n^{0})
+(q-q^{-1})(n^{-}\otimes n^{+}+n^{0}\otimes {\bf 1})~.
\end{array}
\end{equation}
By $\vert\alpha\beta\rangle$ we denote
$\vert\alpha\rangle\otimes\vert\beta\rangle$ for $\alpha,\beta=+,-,0$.
The ferromagnetic states
$\Psi_{\uparrow\uparrow}=\vert\uparrow\uparrow\rangle$ and
$\Psi_{\downarrow\downarrow}=\vert\downarrow\downarrow\rangle$ are
eigenstates of $H$ with energy $q$. The two-hole state
$\Psi_{00}=\vert 00\rangle$ is also an eigenstate with energy $-q^{-1}$.
For configuration of one spin up electron and one spin down electron, the
state is of the form
$\Psi_{\uparrow\downarrow}=a\vert\uparrow\downarrow\rangle
+b\vert\downarrow\uparrow\rangle$. $a$ and $b$ are constants. The
Schr\"odinger equation
\begin{equation}\label{he}
H\Psi_{\uparrow\downarrow}=E\Psi
\end{equation}
gives two solutions of
$E$, $E_{1}=q$, $E_{2}=-q^{-1}$ with eigenstates
\begin{equation}
\begin{array}{l}
\Psi_{\uparrow\downarrow}^{1}=q\vert\downarrow\uparrow\rangle+
\vert\uparrow\downarrow\rangle~,\\[3mm]
\Psi_{\uparrow\downarrow}^{2}=\vert\downarrow\uparrow\rangle-
q\vert\uparrow\downarrow\rangle~,
\end{array}
\end{equation}
respectively.

For the configuration with a hole and a spin up electron we have
\begin{equation}
\begin{array}{l}
\Psi_{0\uparrow}^{1}=q\vert 0\uparrow\rangle+
\vert\uparrow 0\rangle~,\\[3mm]
\Psi_{0\uparrow}^{2}=\vert 0\uparrow\rangle-
q\vert\uparrow 0\rangle~,
\end{array}
\end{equation}
with energy $E_{1}$ and $E_{2}$ respectively.

The one spin down electron and one hole case also gives two eigenstates
\begin{equation}
\begin{array}{l}
\Psi_{0\downarrow}^{1}=q\vert 0\downarrow\rangle+
\vert\downarrow 0\rangle~,\\[3mm]
\Psi_{0\downarrow}^{2}=\vert 0\downarrow\rangle-q\vert\downarrow 0\rangle~,
\end{array}
\end{equation}
with respect to energy $E_{1}$ and $E_{2}$.

Therefore there $9(=3^{L}=3^{2})$ independent states. According to the tensor
decomposition of the $SU_{q}(1,2)$ algebra representation space, there are two
invariant spaces. The eigenstates in the same invariant space can be
exchanged by using the co-product operators of the algebra. One of the
invariant subspace is constituted of states
$\Psi_{\uparrow\uparrow}$, $\Psi_{\uparrow\downarrow}^{1}$,
$\Psi_{0\uparrow}^{1}$, $\Psi_{0\downarrow}^{1}$ and
$\Psi_{\downarrow\downarrow}$
with energy $q$. They satisfy
the following exchange diagram

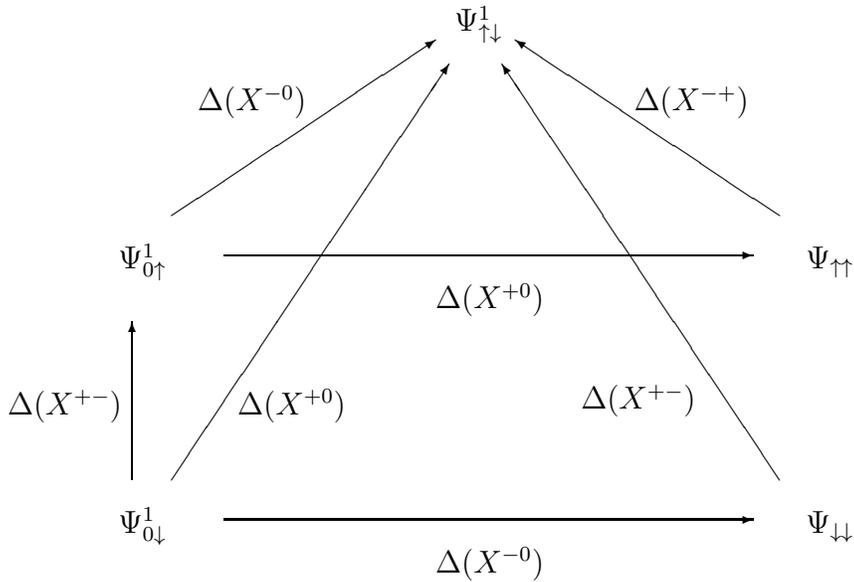
\begin{center}
\begin{picture}(320,240)(0,0)
\put(20,20){$\Psi_{0\downarrow}^{1}$}
\put(280,20){$\Psi_{\downarrow\downarrow}$}
\put(20,120){$\Psi_{0\uparrow}^{1}$}
\put(280,120){$\Psi_{\uparrow\uparrow}$}
\put(147,210){$\Psi_{\uparrow\downarrow}^{1}$}

\put(60,25){\vector(1,0){200}}
\put(60,125){\vector(1,0){200}}
\put(140,5){$\Delta(X^{-0})$}
\put(140,105){$\Delta(X^{+0})$}

\put(25,40){\vector(0,1){60}}
\put(-22,65){$\Delta(X^{+-})$}

\put(40,40){\vector(2,3){105}}
\put(270,40){\vector(-2,3){105}}
\put(65,65){$\Delta(X^{+0})$}
\put(195,67){$\Delta(X^{+-})$}

\put(40,140){\vector(3,2){100}}
\put(270,140){\vector(-3,2){100}}
\put(50,180){$\Delta(X^{-0})$}
\put(215,180){$\Delta(X^{-+})$}
\end{picture}
\end{center}

The left four states give rise to another invariant eigenstate
space of energy $-q^{-1}$,
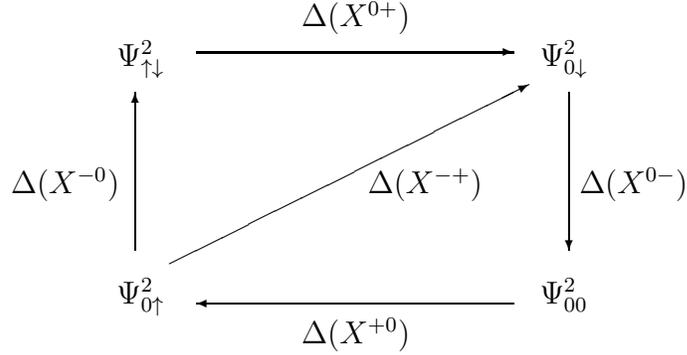
\begin{center}
\begin{picture}(220,140)(0,-10)
\put(20,10){$\Psi_{0\uparrow}^{2}$}
\put(20,100){$\Psi_{\uparrow\downarrow}^{2}$}
\put(180,10){$\Psi_{00}^{2}$}
\put(180,100){$\Psi_{0\downarrow}^{2}$}
%\put(90,25){$\Delta(X^{0+})$}
\put(90,-5){$\Delta(X^{+0})$}
\put(90,115){$\Delta(X^{0+})$}
%\put(90,85){$\Delta(X^{+0})$}
%\put(50,15){\vector(1,0){120}}
\put(170,10){\vector(-1,0){120}}
\put(50,105){\vector(1,0){120}}
%\put(170,100){\vector(-1,0){120}}
%\put(31,90){\vector(0,-1){60}}
\put(27,30){\vector(0,1){60}}
\put(191,90){\vector(0,-1){60}}
%\put(187,30){\vector(0,1){60}}
\put(-20,52){$\Delta(X^{-0})$}
%\put(35,52){$\Delta(X^{0-})$}
%\put(140,52){$\Delta(X^{-0})$}
\put(195,52){$\Delta(X^{0-})$}
\put(40,25){\vector(2,1){136}}
\put(115,52){$\Delta(X^{-+})$}
\end{picture}
\end{center}
Where the inverse action of $\Delta(X^{\alpha\beta})$ is
$\Delta(X^{\beta\alpha})$ for $\alpha\neq\beta=+,-,0$.

For $L=3$ there are $27$ independent eigenstates. They decompose to four
invariant subspaces of the algebra $SU_{q}(1,2)$.

1)~ $E^{\bf 1}=2q$
\begin{equation}
\begin{array}{rcl}
\Psi^{\bf 1}_{1}&=&\vert\uparrow\uparrow\uparrow\rangle\\[3mm]
\Psi^{\bf 1}_{2}&=&q^{2}\vert\downarrow\uparrow
\uparrow\rangle
+ q\vert\uparrow\downarrow\uparrow\rangle
+ \vert\uparrow\uparrow\downarrow\rangle\\[3mm]
\Psi^{\bf 1}_{3}&=&q^{2}\vert 0\uparrow\uparrow\rangle
+ q\vert\uparrow 0\uparrow\rangle
+ \vert\uparrow\uparrow 0\rangle\\[3mm]
\Psi^{\bf 1}_{4}&=&q^{2}\vert\downarrow\downarrow
\uparrow\rangle + q\vert\downarrow\uparrow
\downarrow\rangle
+ \vert\uparrow\downarrow\downarrow\rangle\\[3mm]
\Psi^{\bf 1}_{5}&=&q^{3}\vert 0\downarrow\uparrow
\rangle + q^{2}\vert 0\uparrow\downarrow
\rangle +
q^{2}\vert\downarrow 0\uparrow\rangle\\[3mm]
&&+q \vert\downarrow\uparrow 0\rangle
+ q\vert\uparrow 0\downarrow\rangle
+ \vert\uparrow\downarrow 0\rangle\\[3mm]
\Psi^{\bf 1}_{6}&=&q^{2} \vert 0\downarrow\downarrow
\rangle + q\vert\downarrow 0
\downarrow\rangle + \vert\downarrow\downarrow
 0\rangle\\[3mm]
\Psi^{\bf 1}_{7}&=&\vert\downarrow\downarrow
\downarrow\rangle\\[3mm]
\end{array}
\end{equation}

2)~ $E^{\bf 2}=-2q^{-1}$
\begin{equation}
\begin{array}{rcl}
\Psi^{\bf 2}_{1}&=&\vert 0 0 0\rangle\\[3mm]
\Psi^{\bf 2}_{2}&=&\vert 0 0\downarrow\rangle
- q\vert 0\downarrow 0\rangle
+ q^{2}\vert\downarrow 0 0\rangle\\[3mm]
\Psi^{\bf 2}_{3}&=&\vert 0 0\uparrow\rangle
- q\vert 0\uparrow 0\rangle
+ q^{2}\vert\uparrow 0 0\rangle\\[3mm]
\Psi^{\bf 2}_{4}&=&\vert 0\downarrow\uparrow\rangle
- q\vert 0\uparrow\downarrow\rangle
- q\vert\downarrow 0\uparrow\rangle \\[3mm]
&&+q^{2} \vert\downarrow\uparrow 0\rangle
+ q^{2}\vert\uparrow 0\downarrow\rangle
- q^{3}\vert\uparrow\downarrow 0\rangle
\end{array}
\end{equation}

3)~ $E^{\bf 3}=q-1-q^{-1}$
\begin{equation}
\begin{array}{rcl}
\Psi^{\bf 3}_{1}&=&\vert 0\uparrow\downarrow\rangle
- \vert\downarrow 0\uparrow\rangle
+ \vert\downarrow\uparrow 0\rangle
- \vert\uparrow 0\downarrow\rangle
\\[3mm]
\Psi^{\bf 3}_{2}&=& - \vert\downarrow\downarrow\uparrow
\rangle + q\vert\downarrow\uparrow\downarrow\rangle
+ \vert\downarrow\uparrow\downarrow\rangle
- q\vert\uparrow\downarrow\downarrow\rangle
\\[3mm]
\Psi^{\bf 3}_{3}&=& - \vert\downarrow\uparrow\uparrow
\rangle + q\vert\uparrow\downarrow
\uparrow\rangle
+ \vert\uparrow\downarrow\uparrow\rangle
- q\vert\uparrow\uparrow\downarrow\rangle
\\[3mm]
\Psi^{\bf 3}_{4}&=& - q\vert 0\downarrow\uparrow
\rangle + q\vert 0\uparrow\downarrow
\rangle + q^{2}\vert\downarrow 0
\uparrow\rangle - \vert\downarrow
0\uparrow\rangle\\[3mm]
&&- q^{2}\vert\downarrow\uparrow 0\rangle
+ q\vert\downarrow\uparrow 0\rangle
+ \vert\downarrow\uparrow 0\rangle
- q\vert\uparrow\downarrow 0\rangle\\[3mm]
\Psi^{\bf 3}_{5}&=&- \vert 0\uparrow\uparrow\rangle
+ q\vert\uparrow 0\uparrow\rangle
+ \vert\uparrow 0\uparrow\rangle
- q\vert\uparrow\uparrow 0\rangle\\[3mm]
\Psi^{\bf 3}_{6}&=&- \vert 0\downarrow\downarrow
\rangle + q\vert\downarrow 0\downarrow
\rangle + \vert\downarrow 0\downarrow
\rangle - q\vert\downarrow\downarrow 0
\rangle\\[3mm]
\Psi^{\bf 3}_{7}&=&q\vert 0 0\uparrow\rangle
- q\vert 0\uparrow 0\rangle
+ \vert 0\uparrow 0\rangle
- \vert\uparrow 0 0\rangle\\[3mm]
\Psi^{\bf 3}_{8}&=&q\vert 0 0\downarrow\rangle
- q\vert 0\downarrow 0\rangle
+ \vert 0\downarrow 0\rangle
- \vert\downarrow 0 0\rangle
\end{array}
\end{equation}

4)~ $E^{\bf 4}=q+1-q^{-1}$
\begin{equation}
\begin{array}{rcl}
\Psi^{\bf 4}_{1}&=&\vert 0\uparrow\uparrow\rangle
- q\vert\uparrow 0\uparrow\rangle
+ \vert\uparrow 0\uparrow\rangle
- q\vert\uparrow\uparrow 0\rangle
\\[3mm]
\Psi^{\bf 4}_{2}&=&\vert 0\downarrow\downarrow\rangle
- q\vert\downarrow 0\downarrow\rangle
+ \vert\downarrow 0\downarrow\rangle
- q\vert\downarrow\downarrow 0\rangle\\[3mm]
\Psi^{\bf 4}_{3}&=&q\vert 0 0\uparrow\rangle
+ q\vert 0\uparrow 0\rangle
+ \vert 0\uparrow 0\rangle
+ \vert\uparrow 0 0\rangle
\\[3mm]
\Psi^{\bf 4}_{4}&=&q\vert 0 0\downarrow\rangle
+ q\vert 0\downarrow 0\rangle
+ \vert 0\downarrow 0\rangle
+ \vert\downarrow 0 0\rangle\\[3mm]
\Psi^{\bf 4}_{5}&=&- q\vert 0\downarrow\uparrow
\rangle + q^{2}\vert 0\uparrow\downarrow
\rangle - q\vert 0\uparrow\downarrow
\rangle - \vert 0\uparrow\downarrow\rangle\\[3mm]
&&+ q\vert\downarrow\uparrow 0\rangle
+ q^{2}\vert\uparrow 0\downarrow\rangle
- \vert\uparrow 0\downarrow\rangle
+ q\vert\uparrow\downarrow 0\rangle\\[3mm]
\Psi^{\bf 4}_{6}&=&\vert\downarrow\uparrow\uparrow
\rangle - q\vert\uparrow\downarrow
\uparrow\rangle
+ \vert\uparrow\downarrow\uparrow\rangle
- q\vert\uparrow\uparrow\downarrow\rangle
\\[3mm]
\Psi^{\bf 4}_{7}&=&\vert\downarrow\downarrow\uparrow
\rangle - q\vert\downarrow\uparrow
\downarrow\rangle
+ \vert\downarrow\uparrow\downarrow\rangle
- q\vert\uparrow\downarrow\downarrow\rangle
\\[3mm]
\Psi^{\bf 4}_{8}&=&- \vert 0\uparrow\downarrow\rangle
+ \vert\downarrow 0\uparrow\rangle
+ \vert\downarrow\uparrow 0\rangle
- \vert\uparrow 0\downarrow\rangle
\end{array}
\end{equation}

{}From above calculations of finite lattices it is obvious that the energy
spectrums various with the parameter $q$. For $L=3$ the energy gap between
ground and first excite states is changed from $1$ to $q+q^{-1}-1$. And
when $q$ is taken be minus some energy levels are reversed.

For arbitrary lattice sites we consider the configuration
\begin{equation}
\Psi_{\downarrow}
=\sum_{x=1}^{L}\alpha(x)\vert\uparrow\cdots\uparrow_{x-1}\downarrow_{x}
\uparrow_{x+1}\cdots\uparrow\rangle~.
\end{equation}
The Schr\"odinger equation gives rise to
\begin{equation}\label{deduce}
\begin{array}{l}
E\alpha(x)=q(L-3)\alpha(x)+\alpha(x-1)+\alpha(x+1)+(q-q^{-1})\alpha(x)~,~~~
x\neq 1,L\\[3mm]
E\alpha(1)=q(L-2)\alpha(1)+(q-q^{-1})\alpha(1)+\alpha(2)~,\\[3mm]
E\alpha(L)=q(L-2)\alpha(L)+\alpha(L-1)~.
\end{array}
\end{equation}
This equation set can be solved using the usual Bethe ansatz
\begin{equation}\label{be}
\alpha(x)=A(k)e^{ikx}-A(-k)e^{-ikx}~.
\end{equation}
Substituting (\ref{be}) into the first equation of (\ref{deduce}) one can
get eigenvalues of the Hamiltonian
\begin{equation}\label{e}
E=(L-3)q+(q-q^{-1})+2\cos k~.
\end{equation}
Other equations of (\ref{deduce}) give the ratio of the amplitude $A(k)$
and $A(-k)$
\begin{equation}\label{ration1}
\frac{A(k)}{A(-k)}=\frac{1-qe^{-ik}}{1-qe^{ik}}
\end{equation}
and
\begin{equation}\label{ration2}
\frac{A(k)}{A(-k)}=\frac{(e^{-ik}-q{-1})e^{-ikL}}{(e^{ik}-q{-1})e^{ikL}}~.
\end{equation}
These two equations must be compatible, which gives a constraint on the
impulse $k$,
\begin{equation}\label{k}
e^{2Lk}=1~.
\end{equation}
Therefore
\begin{equation}
k=\frac{l\pi}{L}~,~~~~l=0,1,...,2L-1
\end{equation}
and
\begin{equation}
A(k)=1-q\,e^{-ik}~.
\end{equation}

Now we analyze the spectrum in this special case. First we note that
$k=0$ and $k=\pi$ should be ruled out since they give vanishing Bethe
ansatz function. Second, one can divide the possible impulses $k$ into two
parts, {\bf I}: $(\frac{\pi}{L},\frac{2\pi}{L},\cdots,\frac{(L-1)\pi}{L})$
and {\bf II}:~$(\pi+\frac{\pi}{L},\pi+\frac{2\pi}{L},
\cdots,2\pi-\frac{\pi}{L})$. By
redefining $\tilde{k}_{j}=2\pi-k_{j}$, $k_{j}\in {\bf II}$, one can show
that they give the same Bethe wave functions as $k_{j}\in {\bf I}$.
Therefore, we have only $L-1$ possible impulses and $L-1$ independent wave
vectors. On the other hand, for the configuration with $L-1$ spin-up and
one spin-down electrons, there exit $L$ independent states. While the Bethe
ansatz gives only $L-1$ states. The missing state can be compensated by
using the $SU_{q}(1,2)$ algebra operator ${\cal X}^{-+}$ acting on the all
spin-up state. This state has the same eigenenergy as the all spin-up state
because ${\cal X}^{-+}$ and $H^{q}_{tj}$ are commutative.

{}From this simple special case, we see that the Bethe ansatz equation can
give all the energy spectrum, but not complete states. This had been
pointed out in other integrable models \cite{yue,korepin1}.
It is worthy to note that
Bethe ansatz states are the highest weight states on which the symmetric
group acting gives all complete states. For other configurations, this
conclusion is also true.

For the configuration with one hole and $L-1$ spin-up electrons, we can
similarly write down the wave function,
\begin{equation}
\Psi_{0}
=\sum_{x=1}^{L}\beta(x)\vert\uparrow\cdots\uparrow_{x-1}0_{x}
\uparrow_{x+1}\cdots\uparrow\rangle~.
\end{equation}
The Schr\"odinger equation gives rise to
\begin{equation}\label{deduce0}
\begin{array}{l}
E\beta(x)=q(L-3)\beta(x)-\beta(x-1)-\beta(x+1)~,~~~x\neq 1,L\\[3mm]
E\beta(1)=q(L-2)\beta(1)-\beta(2)+(q-q^{-1})\beta(1)~,\\[3mm]
E\beta(L)=q(L-2)\beta(L)-\beta(L-1)~.
\end{array}
\end{equation}
Using Bethe ansatz it is easy to find
\begin{equation}\label{e0}
E=(L-3)q+(q-q^{-1})+2\cos k~.
\end{equation}
\begin{equation}\label{beta}
\beta(x)=(-1)^{x}\left\{(1-qe^{-ik})e^{ikx}-(1-qe^{ik})e^{-ikx}\right\}
\end{equation}
and
\begin{equation}
k=\frac{\pi}{L},\frac{2\pi}{L},\cdots,\frac{(L-1)\pi}{L}~.
\end{equation}
These give $L-1$ independent states. The missing state can be obtained by
using the operator ${\cal X}^{0+}$ acting on the ferromagnetic states.

One can again see that the symmetry of the system helps in making up the
missing states in Bethe ansatz approach. Hence investigating the symmetry
of integrable system is not an artificial technique. It is the basis in
studying the completeness of the Hilbert space of the system and discussing
related physical properties.
For detailed Bethe ansatz solutions of this generalized t-j model see
\cite{we}.

\vspace{2ex}
\bigskip
\smallskip
{\raggedright{\large \bf Acknowledgement}}
\bigskip

The authors would like to thank Professor P. C. Pu for instructive
discussions. We are also thankful to
CCAST (World Laboratory) for hospitality and support.

\end{document}